\documentclass[[final,3p,times,twocolumn    ]{elsarticle}
\usepackage{graphicx}
\usepackage{dcolumn}
\usepackage{bm}
\usepackage{amsmath}
\usepackage{comment}
\usepackage{epstopdf} 
\usepackage{xr-hyper}
\usepackage{lineno,hyperref}
\usepackage{color} 
\usepackage{soul} 

\makeatletter

\usepackage{amssymb}

\usepackage{lineno,hyperref}
\modulolinenumbers[5]

\journal{Ultramicroscopy}

\begin{document}

\begin{frontmatter}



\title{Phase offset method of ptychographic contrast reversal correction}

\author[emat,nanolab]{Christoph Hofer}
\author[emat,nanolab]{Chuang Gao}
\author[emat,nanolab]{Tamazouzt Chennit}
\author[emat,nanolab]{Biao Yuan}
\author[emat,nanolab]{Timothy J. Pennycook\corref{mycorrespondingauthor}}
\cortext[mycorrespondingauthor]{Corresponding author}
\ead{timothy.pennycook@uantwerp.be}

\address[emat]{EMAT, University of Antwerp, Groenenborgerlaan 171, 2020 Antwerp, Belgium}
\address[nanolab]{NANOlab Center of Excellence, University of Antwerp, Groenenborgerlaan 171, 2020 Antwerp, Belgium}

\begin{abstract}
The contrast transfer function of direct ptychography methods such as the single side band (SSB) method are single signed, yet these methods still sometimes exhibit contrast reversals, most often where the projected potentials are strong. In thicker samples central focusing often provides the best ptychographic contrast as this leads to defocus variations within the sample canceling out. However focusing away from the entrance surface is often undesirable as this degrades the annular dark field (ADF) signal. Here we discuss how phase wrap asymptotes in the frequency response of SSB ptychography give rise to contrast reversals, without the need for dynamical scattering, and how these can be counteracted by manipulating the phases such that the asymptotes are either shifted to higher frequencies or damped via amplitude modulation. This is what enables post collection defocus correction of contrast reversals. However, the phase offset method of counteracting contrast reversals we introduce here is generally found to be superior to post collection application of defocus, with greater reliability and generally stronger contrast. Importantly, the phase offset method also works for thin and thick samples where central focusing does not. Finally, the independence of the method from focus is useful for optical sectioning involving ptychography, improving interpretability by better disentangling the effects of strong potentials and focus.  
\end{abstract}




\end{frontmatter}



\section{Introduction}
Electron ptychography offers very high dose efficiency~\cite{PENNYCOOK2019,zhou2020low_dose,pei2023cryogenic}, the ability to reveal the locations of light elements neighboring heavy atoms~\cite{Yang2016,Gao2022}, post collection aberration correction and superresolution~\cite{Jiang_ptycho,Chen2021}. These advantages make the method a very attractive complement to Z-contrast annular dark field (ADF) workflows~\cite{Krivanek2010}, with the phase images providing stronger images of the structure and the Z-contrast stronger sensitivity to composition. With advances in cameras having greatly reduced or completely removed the problem of slow scans with 4D STEM~\cite{JANNIS2022,Stroppa2023}, there is now relatively little drawback to collecting data for ptychography.

Compared to phase contrast imaging with conventional high resolution transmission electron microscopy (HRTEM), the contrast transfer function (CTF) of direct focused probe methods such as single side band (SSB)~\cite{PENNYCOOK2015,YANG2015,OLEARY2021} and Wigner distribution deconvolution~\cite{YANG2017} are very simple, requiring no aberrations to form contrast and exhibiting no zero crossings. This makes these ptychographic methods much easier to interpret than HRTEM, at least for potentials that are not overly strong. The phase is related to the strength of the potential encountered by the beam electrons and thus as the strength of the potential increases eventually the phase can exceed the limit imposed by the $2\pi$ range of values available to phase and wrap around. This means that as phase increases it can suddenly go from being maximally positive at $\pi$ to maximally negative at -$\pi$, causing a very large change in contrast from only a small change in the sample. Therefore, even though the SSB ptychography CTF, derived using the same weak phase approximation as HRTEM CTFs, shows all frequencies being passed with the same sign, contrast reversals can occur because of wrap around. 


One of the most observed contrast reversal behaviour in atomic resolution imaging is a dip in the phase at the center of the atomic columns~\cite{Clark2022,Gao2022,YANG2017}. These often appear as donut shapes in the images and like a volcano with a caldera in line profiles, and represents a reversal from the centrally peaked probe shaped atoms one observes when the potential is weaker. This makes some intuitive sense as the center of an atomic column in the location of the strongest potential, and thus it is natural to expect that this will be where wrap around will occur first as the potential increases. This intuitive expectation is also in accord with the fact that it is the heavier atomic columns that exhibit donuts first as thickness increases~\cite{Gao2022}. 
Such contrast reversals have also been observed in iCoM and iterative ptychography~\cite{Clark2022,ALLARS21} as well as S-Matrix inversion~\cite{Brown2022,Pelz2021,TerzoudisLumsden2023ResolutionOV}. This again makes intuitive sense as all these methods are attempting to retrieve the same phase shift induced on the beam electrons by the sample. Perhaps less intuitive is the fact that the range of phase values in the final images generally remains much less than $2\pi$ in atomic resolution imaging, at least in single slice ptychography, but it is not just the phases in the final image that can phase wrap – the individual frequency components can also phase wrap. 

If the goal is to locate the light elements hidden in the ADF signal by strong scattering of nearby heavy elements, the appearance of donut contrast on the heavier columns is often not a significant impediment. However as the thickness increases the contrast can become more complex~\cite{Gao2022}. Furthermore, it is often preferable for images of atomic structure to resemble as much as possible to the relatively simple probe shaped spikes in intensity occurring in ADF imaging, even if simply for ease of interpretation. However the contrast reversals can also degrade overall contrast and reduce the visibility of the structure at lower doses~\cite {gao2023central}, as well as complicate quantification.   

For quite a range of thicknesses, central focusing of the beam offers phase images free from contrast reversals and with the strongest contrast overall~\cite{Gao2022,Clark2022,gao2023central}. However this conflicts with the optimal focus for ADF imaging, the entrance surface. Thus optimizing the probe focus for the phase images during acquisition can significantly degrade the quality of simultaneously acquired ADF images, especially as the sample thickness increases and the distance between the entrance surface and optimal focus for ptychography widens. Fortunately post collection adjustments can be applied. The ability of ptychography to adjust aberrations post collection can be leveraged to apply a post collection defocus which can often remove the contrast reversals~\cite{Gao2022}. However the application of post collection defocus also often reduces the overall contrast, even if the atoms all appear ``atom like'' after the contrast reversal correction. Furthermore, in some cases,  post collection defocus does not remove contrast reversals with satisfying results, and indeed in other cases neither does physically focusing the probe during data acquisition ~\cite{Clark2022,gao2023central}. Another interesting approach to overcome contrast reversals is multislice ptychography~\cite{Maiden2012}. Here, the specimen is divided into multiple slices and the phase is solved in each slice separately. Crucially, each slice is thin enough so that phase wraps are avoided within the slices.

Here we delve deeper into the phase wrapping process causing the contrast reversals in direct ptychography, and demonstrate a superior way to counteract them than post collection defocus, which does not rely on focus. As the potential of a single atom is increased, the phases of the spatial frequencies change nonlinearly, with the higher frequencies changing more quickly than the lower frequencies. This means that as the potential is increased, the higher spatial frequencies eventually hit the limit imposed by the 2$\pi$ range of phases available and wrap around. Once a frequency wraps around, its phase tends to contrast very strongly with the frequencies that have not yet wrapped around. Thus these asymptotes in the phase response produce contrast reversals, and they can do so without any dynamical scattering. Applying defocus can roll the phases back around or reduce the amplitude of the wrapped around frequencies sufficiently that the contrast reversals can be removed. However, we find directly adjusting the phases with an offset applied to the zero frequency (DC) phase is generally a superior  method. We show that this method can robustly counteract contrast reversals regardless of the thickness or initial focus of the probe. Although central focusing remains preferable for the absolute best phase contrast in many cases, the phase offset method provides significantly improved contrast compared to defocus correction when a post collection solution is required. Furthermore, the phase offset method can correct contrast reversals in cases where physical defocus cannot satisfactorily do so. The ability of the phase offset to retain contrast is especially important when the sample is fragile and one has a low dose budget. We demonstrate this experimentally at a dose of at 50~e$^-/$\AA$^{2}$ with a thin highly beam sensitive methylammonium (MA)-PbI\textsubscript{3} perovskite solar cell material\cite{li2019_perovskite,MA_perovskite_review} which exhibits contrast reversals that cannot be corrected by defocus at all, whether applied during data collection or after. Furthermore, because the phase offset does not rely on changing focus, it is a useful tool to apply in the context of optical sectioning in which the focus is vital to determining the 3D location of objects. The phase offset allows one to better interpret high resolution ptychographic focal series where the effects of strong potentials and the focal dependence of the contrast reversals are intertwined. The phase offset method can provide a means of disentangling these effects and seeing the objects as clearly as possible in each slice.

\section{Results and Discussion}

\begin{figure}
\includegraphics[width=0.48\textwidth]{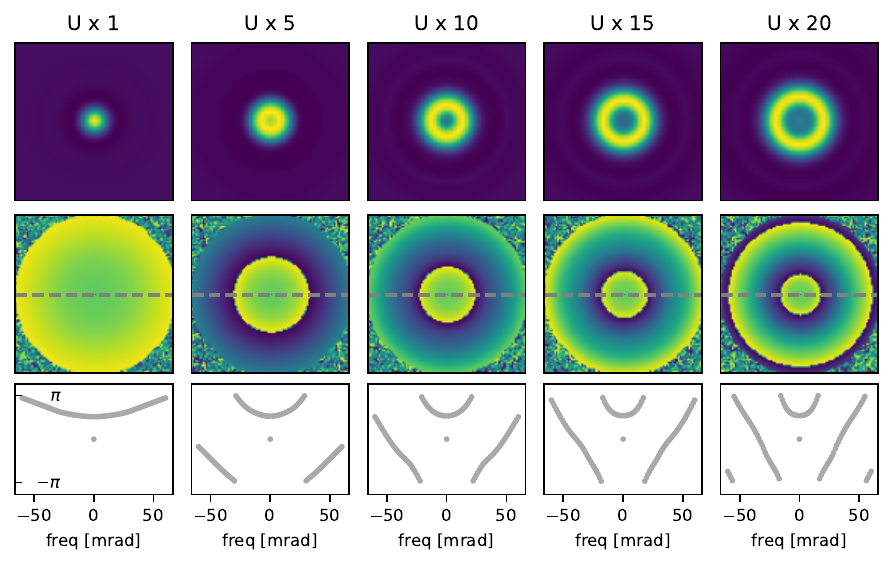}%
\caption{\label{mult}Single atom SSB simulations using potentials ranging from a U atom potential to 20 times that potential,  showing how the potential strength itself causes contrast reversals, which manifest here as donut shaped atom contrast. The 2nd row shows the phase vs frequency in 2D, and the 3rd row line profiles of the rotationally symmetric phase response. As the potential increases the curvature of the phase increases, the phase hits the top limit and wraps around resulting in contrast reversals at the center of the atom. As the wrap around shifts to lower frequencies the donut hole expands. Scale bar is 2~\AA.}
\end{figure}

Figure ~\ref{mult} displays single atom SSB images starting with a U atom and thin increasing its potential incrementally up to a factor of 20. The multiplied potentials correspond to  atoms heavier than any known, but these super heavy atoms allows us to probe the effect of the potential without dynamical scattering complicating matters. 
For SSB reconstructions, we use the  PyPtychoSTEM package~\cite{pyPtychoSTEM} with the 4D data simulated with  abTEM~\cite{abtem} at 200 kV with a 30~mrad convergence angle.

Below the images we plot the phase vs frequency first in 2D, from which it can be seen that phases are continuously rotational symmetric, and then as line plots. The line plots better allow one to see the limits of the 2$\pi$ phase range available. With the already very heavy U atom, the phase curves significantly upwards as a function of the magnitude of the spatial frequency, but remains entirely within the 2$\pi$ phase range without any phase wrap. Multiplying the U potential by a factor of 5, the phase increases more rapidly with respect to frequency, and a phase wrap occurs as the phase extends beyond $\pi$. This results in a significant proportion of the higher spatial frequencies switching from being strongly positive to strongly negative, creating a strong contrast between frequencies lower than and higher than the asymptote, and resulting in a contrast reversal in the form of a small donut hole appearing in the center of the image of the atom. 
As the potential is increased to U $\times 10$, the phase vs frequency curvature further increases. The first wrap around asymptote shifts to lower frequency, a second wrap around point appears and the donut hole increases in size. 
As the potential is further increased the curvature  increases further still, continuing to alter the balance between positive and negative phase frequencies, and further increasing the width of the donut hole. The trend continues as we increase to U $\times 20$, but now the wrapped phase increases in curvature sufficiently to itself surpass the upper limit and itself wrap around.

\begin{figure}
\includegraphics[width=0.48\textwidth]{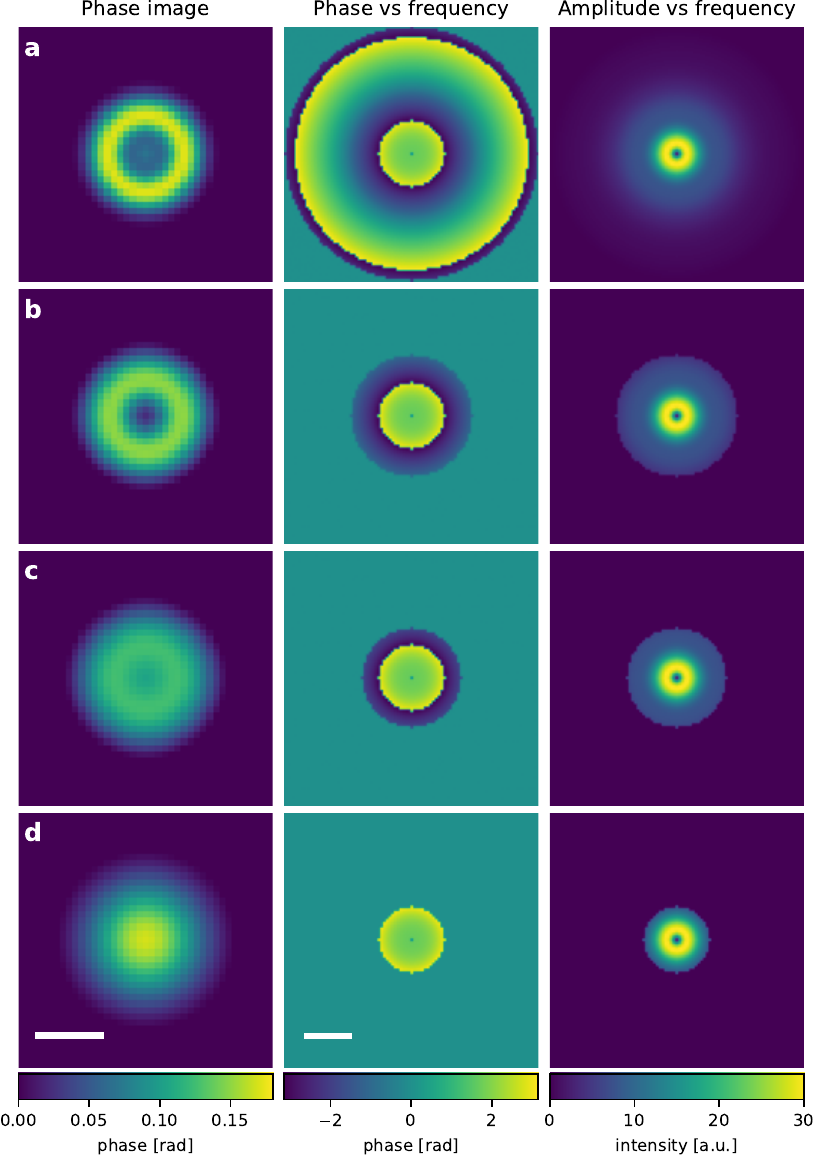}
\caption{\label{mask} Illustration of the effects of the different frequency ranges on the simulated SSB image of the U$\times 20$ potential using masking in Fourier space. With the full range of frequencies out to $2\alpha$ included (a), the atom appears as a thin ring with a slight peak in its center. As we progressively mask out the frequencies after each phase wrap around asymptote (b--d), the positive ring of phase progressively fills more of the central region of the atom, until after removing the contributions from all frequencies above the first wrap around it becomes atom like again, with a peak at the center of the atom. Of course, by limiting the contribution to lower spatial frequencies the image is also limited in resolution. Scale bar is  1~\AA\ for the phase image and 20~mrad for the phase and amplitude vs.\ frequency plot, respectively.  }
\end{figure}

Figure~\ref{mask} shows how the different ranges of frequencies influence the phase image for the U$\times 20$ single atom potential using masking in probe reciprocal space. Before applying any masking, we see the two phase wraps occurring within the $2\alpha$ range of frequencies passed using our 30 mrad convergence angle ($\alpha$), with the last wrap occurring almost at the $2\alpha$ limit of the SSB contrast transfer function~\cite{YANG2015,OLEARY2021}. These two phase wraps define three distinct frequency ranges with strongly contrasting phases. Starting with all spatial frequencies included (Fig.~\ref{mask}a), the single atom appears as a thin ring of strongly positive phase with a lower phase region in the center. The central region has a small peak of slightly more positive phase within the central donut hole with this potential. 

As we mask out the higher frequencies (Fig.~\ref{mask}b) we exclude both the strongly negative frequencies higher than the second phase wrap and the positively phased before it, leaving the negativly phased frequencies after the first phase wrap. T results in the ring of positive phase in the image filling inwards, and the small positive peak in phase at the center of the atom disappears and instead becomes a minimum. As we mask more of the frequencies between the first asymptote and second asymptote, the donut shape fills in more, with the central hole becoming more positive overall but still dipping significantly compared to the phase further from the center of the atom (Fig.~\ref{mask}c). In this step we have further reduced the number of strongly negatively phased frequencies above the first phase wrap asymptote contrasting with the strongly positively phased frequencies below the asymptote. When we remove all the frequencies above the first wrap point, we remove these final strongly contrasting negatively phased frequencies and the atom turns ``atom shaped'' (Fig.~\ref{mask}d).

\begin{figure}
\includegraphics[width=0.48\textwidth]{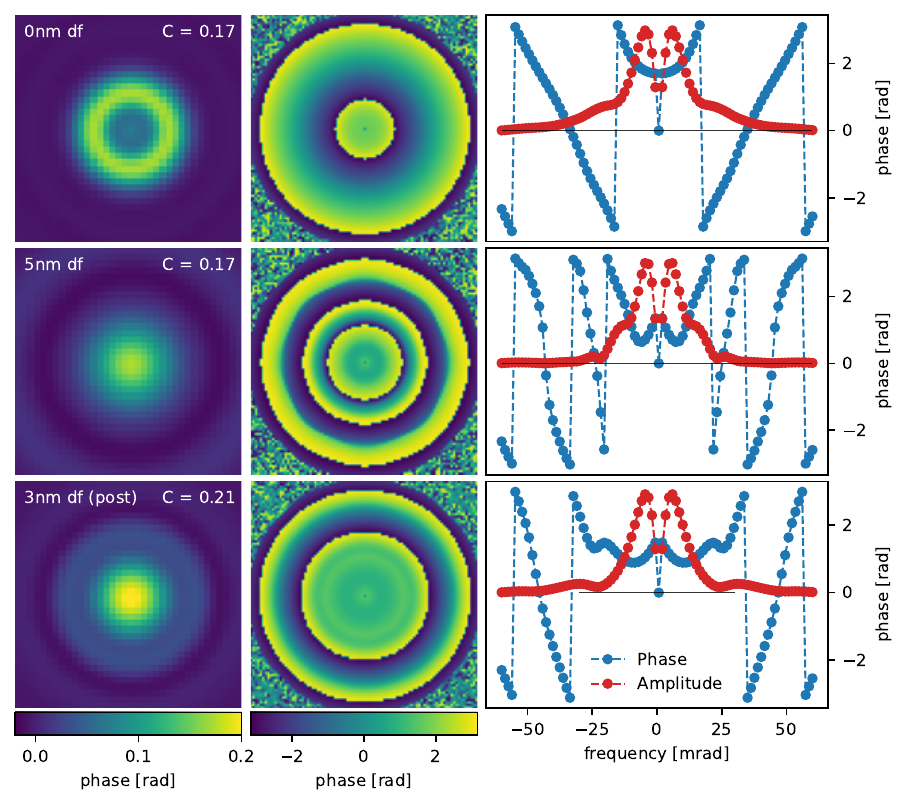}
\caption{\label{wrap} Illustration of the influence of post collection and physical defocus on SSB ptychography with simulations using the U$\times 20$ potential. The SSB images are displayed on the same intensity scale, with the contrast ratio (C) of the maximum phase to background phase indicated. The full frequency response is displayed in the 2D phase vs frequency plots. From the line profiles it is apparent that the physical defocus brings the amplitudes close to zero after the first wrap around, whereas the post collection defocus pushes the first wrap point out to higher spatial frequencies, resulting in stronger contrast in this case.  }
\end{figure}

As has been shown previously, defocus can often be used to counteract contrast reversals~\cite{Gao2022,Clark2022,gao2023central}. For many samples, physically placing the probe focus in the centre of the sample is often found to be optimal.
With a single atom, however, there is no difference between the entrance surface and the center of the sample. Thus there is no significant variation of defocus through the sample, balancing of which being the reason central focusing is optimal in many cases~\cite{gao2023central}. However we can still counteract the contrast reversals of a single atom with physical defocus as shown in figure~\ref{wrap} using a 5~nm defocus with the U$\times 20$ potential. While physical defocus often produces the highest contrast, this is not always the case and the option to counteract the reversals with the probe focused elsewhere can also be a significant benefit. This can often be achieved with the ability of ptychography to alter aberrations post collection~\cite{Gao2022}. Figure~\ref{wrap} illustrates this for the U$\times 20$ potential using a 3~nm defocus applied post collection.

Physical and post collection defocus are generally found to behave somewhat differently, as is also the case here despite this being a single atom with presumably insignificant dynamical effects. Although both physical and post collection defocus remove the donut shape in the SSB image, the behavior further from the atom is different with a ring of higher phase appearing in the post collection defocus results closer to its center than in the physically defocused case. The 2D plot of phase vs frequency is more complex in the physical defocus case here, with the post collection appearing to nonlinearly push the phase wraps out to higher frequency. There are still two phase wraps in the post collection case, but they are concentrated closer to the 2$\alpha$ transfer limit, leaving a broader range of unwraped lower frequencies. In the physically defocused case, it appears more that it is the suppression of the amplitudes of the phases after the first wrap that results in the contrast reversal removal. We observe here three phase wraps, but with low amplitudes the strongly contrasting phases of the wrapped higher frequencies do not contribute significantly to the image.  

However, the reduction of contrast that results from post collection defocus motivated us to search for an alternative strategy to counteract contrast reversals. We present here what we call the phase offset method.
Given that the DC term provides the baseline phase which all other frequencies interfere with when transforming from probe reciprocal space into real space to form an image, by altering the relative phase of the DC term and all the other phases with a rigid offset, we can manipulate the phase wrap point and move it to higher frequencies without otherwise altering the overall shape of the phase vs frequency plot. In practice one can simply shift the DC term itself, although for the purposes of illustration here we instead shift the phases of the other frequencies while keeping the DC term phase constant in our plots of phase vs frequency as this better shows the effect on the phase wraps. 

\begin{figure}
\includegraphics[width=0.48\textwidth]{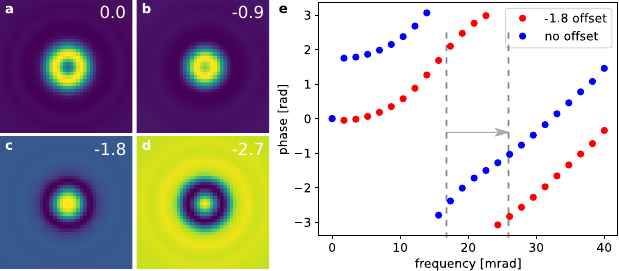}
\caption{\label{U}Illustration of the use of a phase offset on strong phase objects. (a--d) SSB images simulated with the U$\times 10$ potential using phase offsets  of 0.0, -0.9, -1.8 and -2.7 mrad on the nonzero spatial frequencies with the DC values set to zero. (e) Line profiles of the phases with no offset and the optimal -1.8 rad offset, again with only the upper phase vs frequency curve shown for simplicity. }
\end{figure}

Figure ~\ref{U} illustrates the use of a phase offset with the U$\times10$ potential. Without any correction the phase vs frequency curve displays a sharp jump from the DC term to the first nonzero frequency. Applying an offset of -1.8 radians to the nonzero frequency components rigidly shifts the phase curve down such that there is no discontinuity moving from the DC term to the higher spatial frequencies until the positive $\pi$ upper limit is hit and the phase wrap occurs as shown in Figure ~\ref{U}e. Importantly, with this offset the phase wrap occurs at a significantly higher spatial frequency than without the offset, and as can be seen from the figure the resulting image, Figure ~\ref{U}c, is donut free and much more closely resembles the shape of a lighter single atom that does not cause wrap around. There is a negative halo, but this is normal for a single atom in ptychographic images~\cite{HOFER2023kernel,OLEARY2021}. If we increase the phase offset to -2.7 radians the phases instead obtain a negative sign in the low frequency region which contrasts with the now positive higher frequencies and  magnifies the negative halo leading to an ``inverted donut" (cf.~Figure ~\ref{U}d). This implies that the ``best'' offset value is the one which provides as much of a single signed phase curve as possible. 


\begin{figure}
\includegraphics[width=0.48\textwidth]{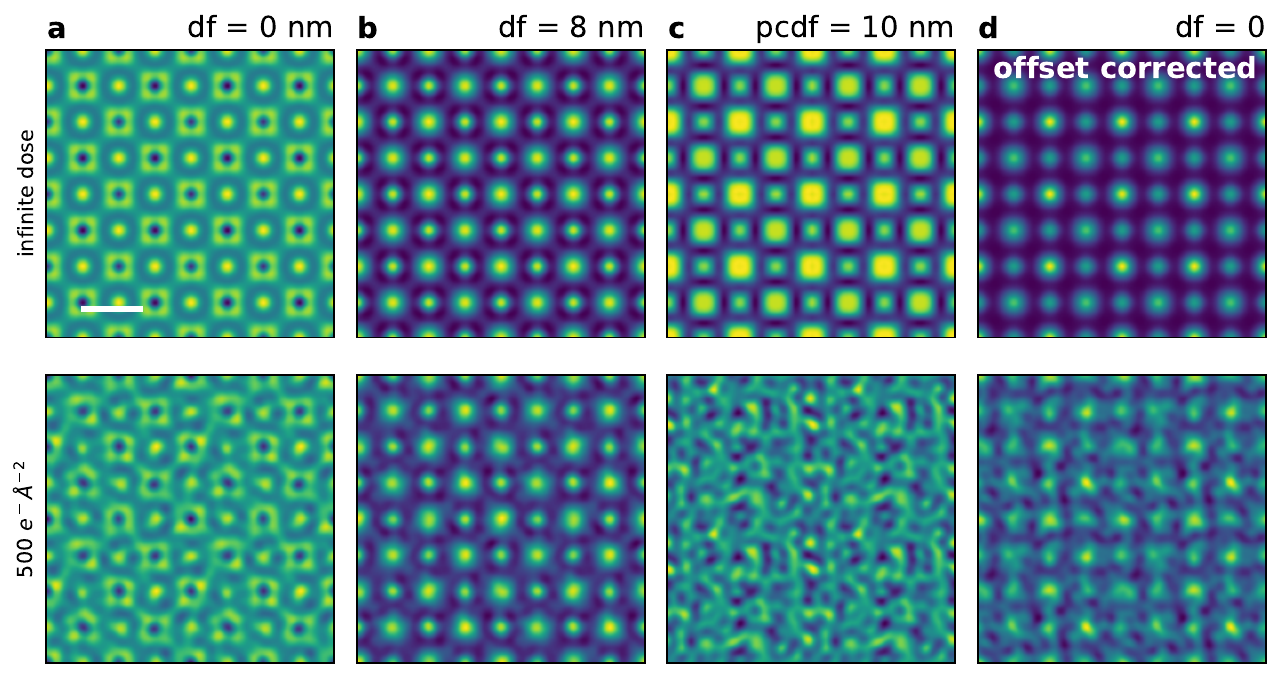}
\caption{\label{comparison}Comparison of SSB images simulated for 16~nm thick STO with the probe focused to the entrance surface (df = 0), the central slice (df = 8 nm) without further correction, and focused to the entrance surface with a post-collection defocus of 10~nm and using the phase offset method. The top row of images is noise free, while the bottom row of images uses a dose of 500~e$^-/$\AA$^{2}$. Here central focusing is optimal, correcting the reversals with strong contrast. The post collection defocus correction is very noisy in the low dose simulation, but the phase offset method retains sufficient contrast to locate all the columns at low dose while retaining the optimal probe focus for the ADF. Scale bar is 3~\AA. }
\end{figure}

While a single atom is a rather simple system, the phase offset method also works well with crystals. As a first example, Fig.~\ref{comparison} examines the use of the phase offset with 16 nm thick SrTiO$_3$ (STO). In panel a the probe is focused to the entrance surface, which we emphasize is the best condition for ADF imaging. However, this leads to contrast reversals at the heavy Sr and Ti sites in the SSB image. Physically focusing to the middle of the specimen, as in Fig.~\ref{comparison}b, removes the contrast reversals as a result of the defocus phase compensation of different layers~\cite{gao2023central}. To correct the contrast reversals using post collection defocus applied to data taken with the probe physically focused to the entrance surface during acquisition requires, in this case, a significantly larger defocus which, as we will show, results in a significant contrast reduction of the phase image. Post collection defocus adjustment often leads to sufficiently large contrast reduction that the atoms are not visible at low doses such as the 500~e$^-/$\AA$^{2}$ used in the bottom row of Fig.~\ref{comparison}. As seen in the figure, the image in which the contrast reversals have been corrected using physical defocus remains quite clear at this dose. Compared to post collection defocus, the phase offset does not reduce the contrast nearly as much, providing an image in which all the locations of the atoms are easily identified also at the lower dose. 

We note that as the ptychographic contrast is not as high here with the focus set at the entrance surface even with the phase offset method, compared to physically focusing to the center of the sample, one must choose to prioritize either optimal ptychographic contrast at the expense of the ADF or having a better ADF contrast by focusing to the entrance surface and compensating the ptychography with a phase offset. Many materials science samples can handle many orders of magnitude higher doses than 500~e$^-/$\AA$^{2}$, and for these one may wish to to optimize the ADF by focusing to the entrance surface while still obtaining a high quality contrast reversal free ptychographic image via the phase offset method. On the other hand, if the dose budget for a given sample is very low, one might consider that one might not obtain useful information from the far less dose efficient ADF signal even with the focus at the entrance surface, and choose to physically focus to the center of the sample. Of course, optimizing the focus under low dose conditions is also very difficult and thus there likely remains benefit to optimizing via post collection adjustments such as the offset method at low doses as well, even if a central focus was the aim.

\begin{figure}
\includegraphics[width=0.48\textwidth]{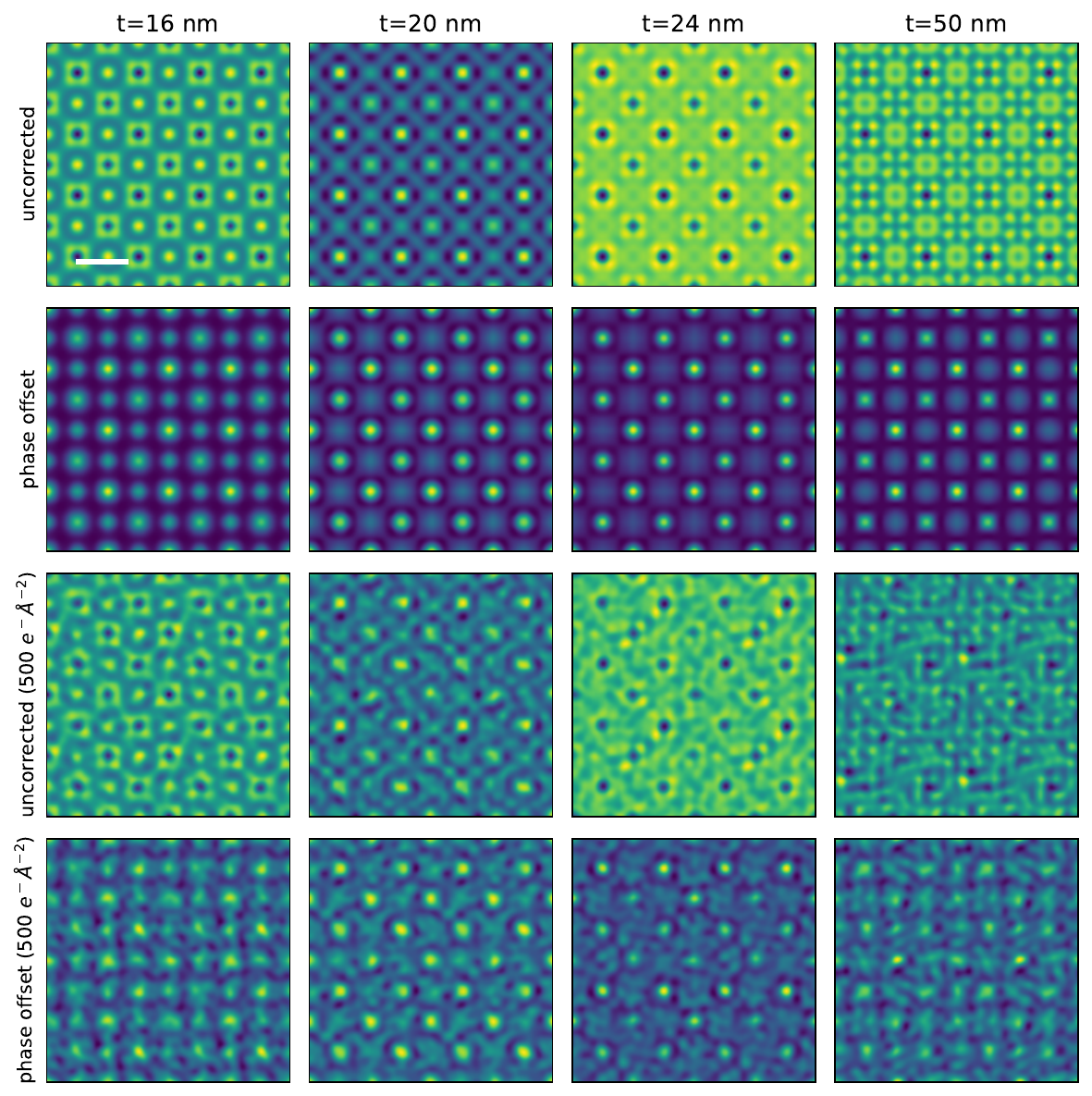}
\caption{\label{thickness} Simulated SSB imaging of STO as a function of thickness (t, indicated above each column) comparing the uncorrected and phase offset results with the probe focused to the entrance surface. The top half of the figure is with infinite dose, and the bottom with a dose of 500~e$^-/$\AA$^{2}$. These results show the robustness of the phase offset method across a wide range of thickness. Scale bar is 3~\AA.}
\end{figure}

\begin{figure}
\centering
\includegraphics[width=0.48\textwidth]{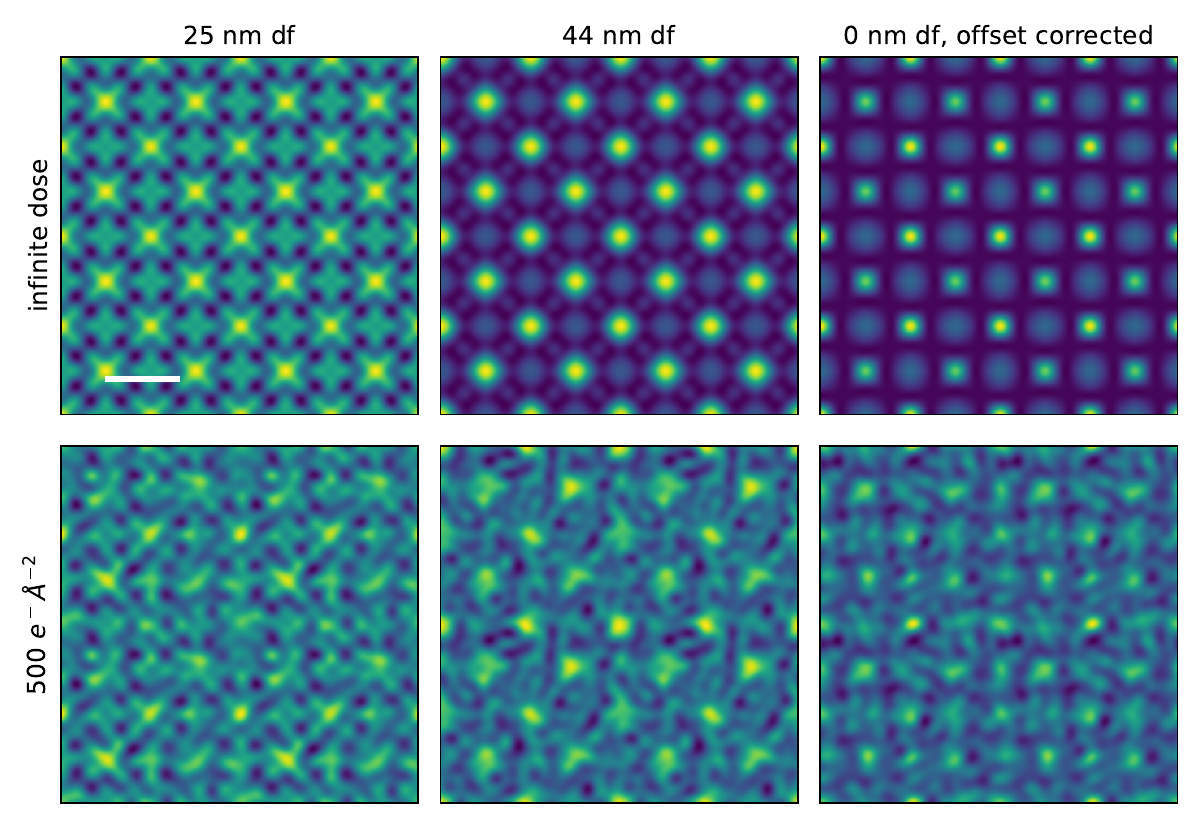}
\caption{\label{50nm} Comparison of physical defocus and the phase offset method for contrast reversal removal with 50 nm thick STO simulations. Central focusing (25 nm defocus) does not remove the contrast reversals in this case. Instead, close to the exit surface, using 44 nm of defocus, was found to be optimal using a focal series. However, physically focusing just 6 nm from the exit surface results in additional atom like features appearing where there are no atoms. This does not occur using the phase offset on data taken with the probe focused to the entrance surface. Furthermore the O columns are more visible at the low 500~e$^-/$\AA$^{2}$ dose using the phase offset method than physical focusing. Scale bar is 3~\AA.}
\end{figure}

Since ptychographic contrast is quite sensitive to the sample thickness, we now demonstrate that the offset correction can be successfully applied to a large variety of thicknesses. Fig.~\ref{thickness} shows STO SSB images noise free and with a dose of 500~e$^-/$\AA$^{2}$ at thicknesses of 16~nm, 20~nm, 24~nm and 50~nm. This covers a range of specimen thicknesses typical of atomic resolution electron microscopy in materials science. Focusing to the entrance surface leads to contrast reversals as seen in the first and third rows of Fig.~\ref{thickness}. The phase offset correction leads to a reasonable contrast with all thicknesses. In the 24~nm case, the oxygen columns have a much weaker contrast in both the uncorrected and the corrected phase images. This can be improved by physically focusing to the center of the sample thickness, as we showed previously~\cite{gao2023central}, however, this can be difficult in experiments in practice without live ptychography. 
For 50~nm of STO, the contrast reversals are sufficiently complex in the uncorrected image that it is practically uninterpretable at 500~e$^-/$\AA$^{2}$. This is a very low dose for STO, but it is nevertheless informative regarding the contrast generally as well for samples that handle only very low doses.

50~nm is also an interesting case because the center of the sample is not the focal plane exhibiting optimal contrast with physical focusing, as we found previously by performing a simulated focal series~\cite{Gao2022}. Figure \ref{50nm} shows that the central focal plane exhibits quite strong contrast complexity that is not ``atom like''. Instead, it was found that physically focusing to near the exit surface provides much better correction of contrast reversals. An example of this is shown in figure \ref{50nm} using a 44 nm defocus from the entrance surface, just 6~nm from the exit surface of the sample. Here the contrast is much better, appearing more atom like but also clearer, including at 500~e$^-/$\AA$^{2}$. However some artifacts remain in the form of atom like spots in between the actual atoms, as is seen in the noise free image. These are not present in the phase offset images, which not only show no contrast reversals or artifacts but actually show more visible contrast on the O sites at 500~e$^-/$\AA$^{2}$. Given the difficulty of optimizing focus during low dose work, the performance of the phase offset here is encouraging.  

\begin{figure}
\includegraphics[width=0.48\textwidth]{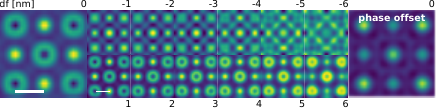}
\caption{\label{df} Simulated focal series for thin 5~nm thick STO showing that no defocus value can counteract the contrast reversals within a range that does not overly distort the images. Importantly, the phase offset method counteracts the contrast reversals with the probe focused to the entrance surface and retains good contrast. Scale bar is 2~\AA.}
\end{figure}

Furthermore, focus adjustment of the beam cannot not always remove contrast reversals. While one might expect that contrast reversals arise only in relatively thick materials, surprisingly thin materials can also exhibit contrast reversals, and these can be impossible to remove with defocus. Clark et al.\ showed this for very thin GaN~\cite{Clark2022}. We have explored this for 5~nm thick STO~\cite{gao2023central}, which we find also exhibits contrast reversals which cannot be counteracted with central focusing. This is perhaps in some sense intuitive given the small range of defocus that exists within the sample. However, the reversals also can not be corrected with any focal point within the sample, or even within a useful range beyond as shown in Fig.~\ref{df}. While perhaps the contrast reversal begins to be counteracted far beyond the exit surface, the contrast has reduced to the point that the O sites are almost invisible. However, if we instead apply a phase offset to the entrance surface focused data the contrast reversals are completely removed. This shows that the phase offset method can in some cases work better than any type of defocus adjustment.

\begin{figure}
\includegraphics[width=0.48\textwidth]{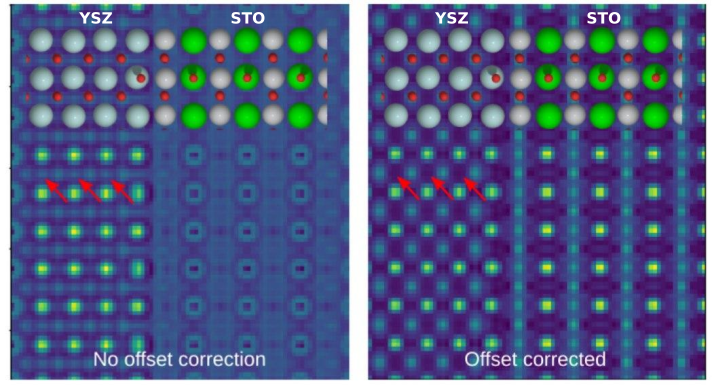}
\caption{\label{interface} Simulated SSB image of a STO/zirconia interface before  and after phase offset correction. Contrast reversals as well as the ``virtual atoms" indicated by the red arrows are removed in the phase offset corrected image.}
\end{figure}

Extending the complexity of the system beyond a bulk crystal, we also tested an oxide heterostructure. Here we again used STO but now interfaced with ZrO$_2$ in the cubic fluorite structure epitaxially lattice matched to the STO. The structure is 16 nm thick in the beam direction. Interestingly, the uncorrected image in Fig.~\ref{interface} (left) shows strong contrast reversals at all STO sites. The zirconia also shows virtual atoms as indicated by the red arrows. All these artifacts are removed by applying an offset. Therefore the phase offset method appears robust to handling more complex structures than pristine bulk crystal structures. The results show again how such contrast reversal correction can be vital to providing meaningful and interpretable images of atomic structures. We therefore conclude that the phase offset correction seems very useful for ptychography of a wide range of materials.

We now demonstrate the phase offset method with experimental data. 4D STEM data of a methylammonium(MA)-PbI\textsubscript{3} perovskite was acquired using our Timepix3 event driven camera to easily achieve very low doses and avoid drift~\cite{JANNIS2022}. A 13~mrad convergence angle was used to optimize for contrast in the frequency range of interest. Due to the extreme beam sensitivity of the material, we use a dose of just 50~e$^-/$\AA$^{2}$. We note that this in the dose regime used in cyro electron microscopy of proteins. Although the event driven camera makes such low dose scans easy to achieve, the very low dose still makes it very difficult to find the best focus during the experiment, especially as one wishes to spend all the dose budget on imaging the regions of interest, not on adjusting the focus. In practice, at present focusing is performed by optimizing the ADF image, which again is often not the optimal defocus for the contrast of the ptychography. 

\begin{figure}
\includegraphics[width=0.48\textwidth]{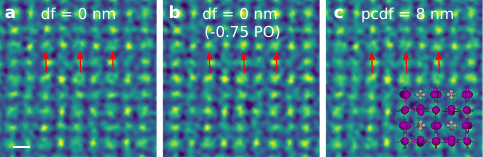}%
\caption{\label{exp} Experimental SSB images of thin ca.\ 2–4~nm thick MAPbI\textsubscript{3}. (a) Uncorrected SSB image taken with the focus optimized for the ADF (not shown). Contrast reversals are clearly visible on the heavy columns as highlighted by the red arrows. The phase offset corrected SSB image (b) removes the contrast reversals while the post collection defocus (pcdf) cannot completely remove the contrast reversals. Scale bar is 3~\AA.}
\end{figure}
An SSB image of the MAPbI\textsubscript{3} is shown without correction in Fig.~\ref{exp}a. The heavy columns, which include Pb, tend to be donut shaped, despite the very low thickness of approximately 4~nm as indicated by electron energy loss spectroscopy. This thickness falls within the regime where the contrast reversals cannot be corrected by defocusing, assuming the STO results are representative as we expect. Indeed a post acquisition defocus series does not remove the contrast reversals without losing so much contrast as to make the atoms essentially invisible. In Fig.~\ref{exp}c shows the result of using an 8 nm post collection defocus which retains sufficient contrast to resolve the atoms but only reduces the contrast reversals rather than removing them. Applying a phase offset, however, completely removes the contrast reversals without any obvious compromise of the contrast, as shown in Fig.~\ref{exp}b. 

The fact that increasing the post collection defocus reduces the contrast to the point of losing the lattice contrast completely is in agreement with the earlier discussion regarding the thin STO and the defocus series shown in Fig.~\ref{df}, where a high defocus only corrects the reversals to a small extent. For low dose data, such as that of the MAPbI\textsubscript{3}, such high defocus values lead to a complete loss of the atomic resolution signal as a result of the contrast reduction associated with defocusing. In this case, the phase offset is the only method that can practically be used to obtain an easily interpretable image without contrast reversals.

\begin{figure}
\centering
\includegraphics[width=0.40\textwidth]{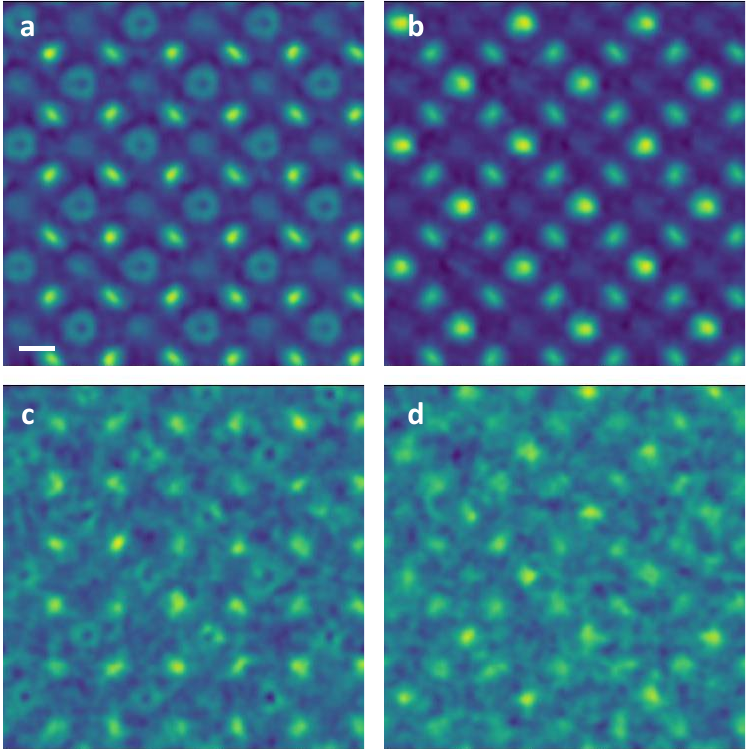}%
\caption{\label{epie} Simulated rPIE images of thin 2.4~nm thick MAPbI\textsubscript{3}. (a) Noise-free uncorrected rPIE image taken with a focused probe. Donuts are visible on the Pb sites, similar to the SSB image. (b) Phase offset corrected version of (a). (c) is the same as (b) but with an electron dose of 1000~e$^-/$\AA$^{2}$ and using a defocus of 50 nm. (d) is the phase offset corrected version of (c) using the same offset as (b). The scale bar is 3~\AA.}
\end{figure}

Since contrast reversals have also been observed in iterative ptychography reconstructions ~\cite{Clark2022}, it is interesting to see if the phase offset can also be used for these methods as one would expect. For this reason, we simulated a 4D data set of MAPbI\textsubscript{3} with a focused probe and processed it with the regularized iterative ptychography (rPIE) algorithm as implemented in abTEM~\cite{abtem}. Donuts appear at the Pb sites, similar to the SSB case, as shown in Fig.~\ref{epie}a. Applying the phase offset to the rPIE result indeed removes the contrast reversals as shown in panel b. A 1000~e$^-/$\AA$^{2}$ dose using a 50 nm defocus version of the phase offset corrected rPIE reconstruction is shown in Fig.~\ref{epie}c. 

\begin{figure}
\centering
\includegraphics[width=0.46\textwidth]{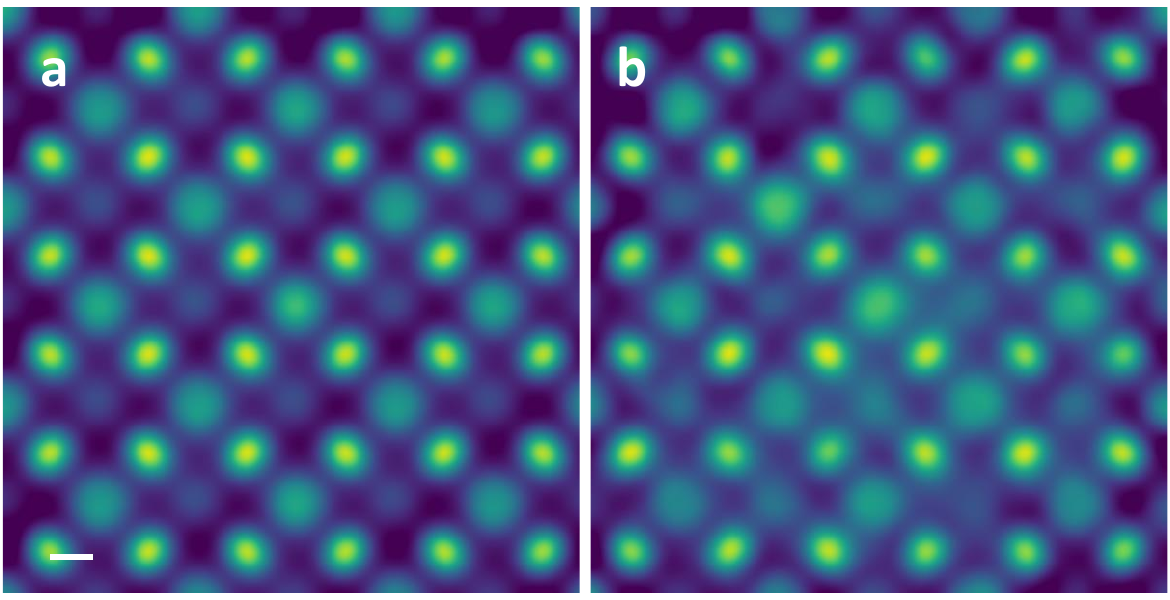}%
\caption{\label{GD} Simulated gradient descent iterative ptychography images of 2.4~nm thick MAPbI\textsubscript{3}. (a) Noise-free reconstruction using a focused probe, with no donuts visible on the Pb sites. (b) is the same as (a) but using a dose of 1000~e$^-/$\AA$^{2}$ no contrast reversals are visible on the reconstruction. Scale bar is 3~\AA.}
\end{figure}

We note that another possible way to avoid contrast reversals due to phase wrapping in reciprocal space in iterative ptychography is to solve for the potential directly rather than the complex object as proposed by~\cite{py4dstem-optica}. We have tested this with the gradient descent algorithm as implemented in the py4DSTEM code, and find that in the case of the thin MAPbI\textsubscript{3} it does indeed also remove the contrast reversals seen with rPIE as shown in Fig.~\ref{GD}. However, we note that the images also appear considerably less sharp.

Optical sectioning makes use of finite depth of field to locate objects in 3D by observing at which focus objects appear most sharp. Optical sectioning with direct ptychography was demonstrated at relatively low resolution by Yang et al. \cite{Yang2016}, using a single scan and altering the defocus post-collection. This allowed the 3D locations of nanotubes to be discerned, and it was shown that the method provides a true optical sectioning effect rather than a Fresnel propagated version of the exit wave. Defocus away from the plane at which an object is located acts to introduce a variation in the phase in the double overlap regions that diminishes the overall amplitude the more an object is out of focus. This is the reason why using defocus to correct contrast reversals overall tends to result in lower contrast as opposed to using the phase offset which does not introduce phase oscillations that diminish the amplitudes. 

For optical sectioning we can therefore use the phase offset without reducing the optical sectioning effect, and by correcting contrast reversals independently of the defocus significantly clean up our view of a sample at different depths, aiding interpretation. Introducing defocus can itself introduce contrast reversals~\cite{Gao2022}, therefore it is valuable to be able to counteract these with another independent method such as the phase offset. 
This is important because interpreting atomic resolution optical sectioning in the presence of contrast reversals can be rather confusing. 

\begin{figure}
\centering
\includegraphics[width=0.46\textwidth]{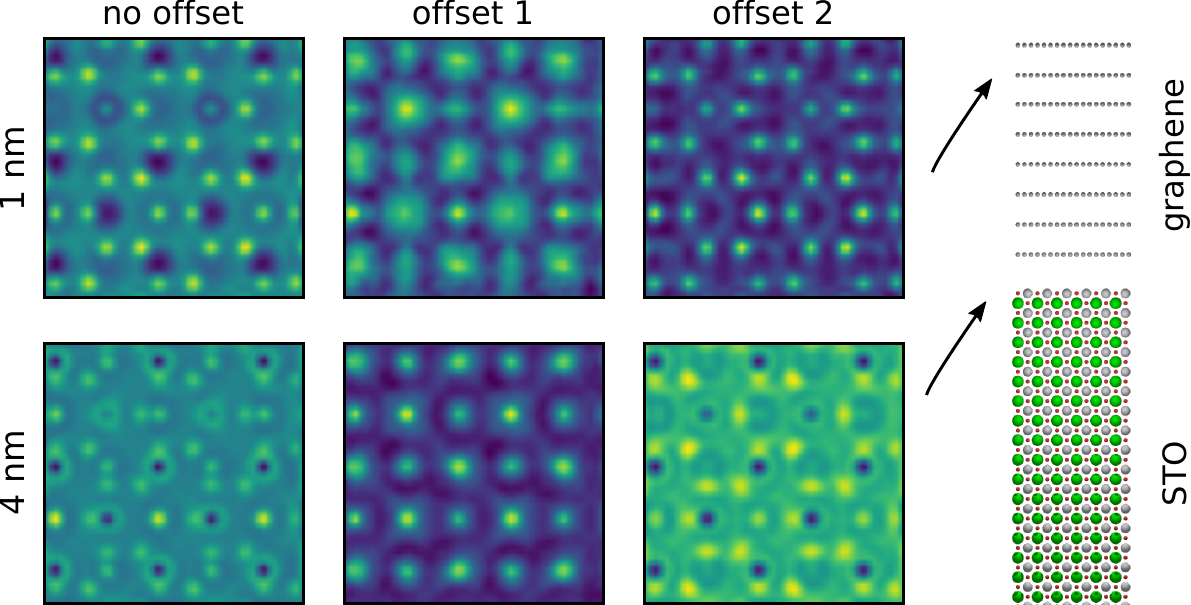}%
\caption{\label{heterostructure} Simulated SSB phase images with the probe focused to two different points in a graphene/STO heterostructure. Top: Focal point 1 nm below the top graphene layer. Bottom: Focal point close to the STO interface. In both cases, the uncorrected image shows a mixture of both structures as well as contrast reversals on the heavy sites of the STO. Offset one maximizes the contrast of the STO while offset two maximizes the contrast of the graphene.}
\end{figure}

To illustrate this we show simulated optical sectioning of 5 nm of graphene placed on top of STO in figure~\ref{heterostructure} using a 30 mrad probe convergence angle. We can optimize the phase offset for viewing either lattice as clearly as possible within the bounds of the optical sectioning effect and otherwise available contrast. Thus for example we can use the phase offset to actually see the STO go out of focus much more clearly as the probe moves into the graphene out of the STO using offset number one compared to using no offset. Offset number two optimizes the contrast of the graphene layers. The C potential is much weaker than the far heavier Sr and Ti columns which are also much denser than the C columns in the widely spaced van der Waals layers of graphene. Therefore it is harder to see the graphene over the out of focus STO, but offset number two still mitigates the dark dips in phase due to the STO inside the graphene.

In conclusion, the phase offset method offers a significant boost to our ability to counteract contrast reversals in electron ptychography. Optimizing the focus used with the data acquisition often provides the best ptychographic contrast, such as central focusing with intermediate thicknesses. However there are many situations where using a focus optimal for the ptychographic contrast is not practical, and indeed cases where defocus cannot be used to counteract contrast reversals at all. Often one may prefer to optimize defocus for the ADF, as this cannot be corrected post collection. At doses sufficiently high that the less efficient ADF signal shows good contrast, the contrast reduction from using the offset method vs a physical defocus optimised for the ptychography will often not be so significant as to matter for locating atoms. At very low doses where the ADF provides very poor contrast, one may choose to abandon the ADF and prioritize focusing for the ptychography. However, the ADF can provide very informative information at surprisingly low doses, even if exceedingly noisy, and in practice accurate focusing is particularly challenging at extremely low doses. Furthermore, defocus cannot always correct contrast reversals as is the case for the thin few nm thick samples we discussed. Thus, as we see with the experimental example with MAPbI\textsubscript{3} here, the phase offset can be an important tool to remove contrast reversals, even at the extremely low doses used in cryo electron microscopy of proteins. Overall, we find the phase offset method reliably overcomes contrast reversals with minimal contrast reduction, or even improved contrast, compared to a defocus optimized for ptychography at the time of acquisition and thus we expect it to become a standard tool in the use of direct electron ptychography. Finally, we find that the offset method can be a useful tool for optical sectioning with ptychography due to its ability to improve contrast and interpretability independent of the focus.

\section{Acknowledgement}
We acknowledge funding from the European Research Council (ERC) under the European Union's Horizon 2020 Research and Innovation Programme via Grant Agreement No. 802123-HDEM (C.H., C.G., T.C., B.Y. and T.J.P.) and FWO Project G013122N “Advancing 4D STEM for atomic scale structure property correlation in 2D materials” (C.H.). 


\end{document}